\documentclass[final,authoryear,1p,times]{elsarticle}

\usepackage{graphicx}

\usepackage{amsmath,amsfonts,amssymb,url}

\renewcommand{\url}{\begingroup \small \Url}

\newcommand{\be}{\begin{equation}}
\newcommand{\ee}{\end{equation}}
\newcommand{\bea}{\begin{eqnarray}}
\newcommand{\eea}{\end{eqnarray}}

\newcommand{\dd}{\text{d}} 

\newcommand{\Ol}{\mathcal{O}}

\newcommand{\C}{\mathcal{C}}

\usepackage[rightcaption]{sidecap}
\bibpunct{(}{)}{;}{a}{}{,}

\newcommand{\beq}{\begin{equation}}
\newcommand{\eq}{\end{equation}}
\newcommand{\bed}{\begin{displaymath}}
\newcommand{\ed}{\end{displaymath}}
\newcommand{\reff}[1]{\mbox{Fig.\hspace{2pt}\ref{#1}}}

\newcommand{\refe}[1]{(\ref{#1})}
\newcommand{\refee}[2]{(\mbox{\ref{#1})\hspace{2pt}--\hspace{2pt}(\ref{#2})}}

\newcommand{\refs}[1]{\mbox{Sec.\hspace{2pt}\ref{#1}}}

\newcommand{\eksc}{\ensuremath{\varepsilon}}
\newcommand{\rp}{r_{\mathrm p}}
\newcommand{\ra}{r_{\mathrm a}}
\newcommand{\rg}{r_{\mathrm{g}}}
\newcommand{\degr}{^{\circ}}
\newcommand{\incl}{\ensuremath{\iota}}

\newcommand{\pp}{\mathcal{P}}

\newcommand{\cn}{\textrm{cn}}

\newcommand{\elF}{\textrm{F}}
\newcommand{\elK}{\textrm{K}}

\begin{document}

\begin{frontmatter}



\title{Numerical modeling of a Global Navigation Satellite System in a general relativistic framework}


\author[label1]{Pac\^ome Delva\corref{cor1}}

\address[label1]{ESA Advanced Concepts Team, ESTEC, DG-PI, Keplerlaan 1, 2201 AZ Noordwijk, The Netherlands}
\cortext[cor1]{Corresponding author}
\ead{pacome.delva@esa.int}

\author[label2]{Uro\v{s} Kosti\'{c}}
\ead{uros.kostic@fmf.uni-lj.si}

\author[label2]{Andrej \v{C}ade\v{z}}

\address[label2]{Department of Physics, University of Ljubljana, Jadranska 19, 1000 Ljubljana, Slovenia}
\ead{andrej.cadez@fmf.uni-lj.si}

\begin{abstract}
In this article we model a Global Navigation Satellite System (GNSS) in a Schwarzschild space-time, as a first approximation of the relativistic geometry around the Earth. The closed time-like and scattering light-like geodesics are obtained analytically, describing respectively trajectories of satellites and electromagnetic signals. We implement an algorithm to calculate Schwarzschild coordinates of a GNSS user who receives proper times sent by four satellites, knowing their orbital parameters; the inverse procedure is implemented to check for consistency. The constellation of satellites therefore realizes a geocentric inertial reference system with no \emph{a priori} realization of a terrestrial reference frame. We show that the calculation is very fast and could be implemented in a real GNSS, as an alternative to usual post-Newtonian corrections. Effects of non-gravitational perturbations on positioning errors are assessed, and methods to reduce them are sketched. In particular, inter-links between satellites could greatly enhance stability and accuracy of the positioning system.
\end{abstract}

\begin{keyword}
General Relativity \sep Schwarzschild Space-time \sep Relativistic Positioning System \sep GNSS

\end{keyword}

\end{frontmatter}



%
\section{Introduction}
\label{s:intro}
The classical concept of positioning system for a Global Navigation Satellite System (GNSS) would work ideally if all satellites and the receiver were at rest in an inertial reference frame. But at the level of precision needed by a GNSS, one has to consider curvature and relativistic inertial effects of space-time, which are far from being negligible. There are two very different ways of including relativity in a positioning system: one way is to keep the newtonian conception of absolute time and space, and add a number of post-newtonian corrections depending on the desired accuracy; another way is to use a relativistic positioning system. This is a complete change of paradigm, as the constellation of satellites is described in a general relativistic framework. This new scheme for positioning potentially allows the definition of a very stable and accurate primary reference system.

For a detailed review of the post-newtonian scheme of a GNSS one can read~\citet{ashby03} and \citet{pascualsanchez07a}. In this scheme, the two main corrections come from gravitational frequency shift between the clocks - due to the local position invariance principle - and from the Doppler shift of the second order - due to relative motion of satellites and users. A ``GPS coordinate time'' is defined as the \emph{time of a clock at rest on the geoid}. To add the post-newtonian corrections, it has to be related to the time measured by a clock in an inertial frame at spatial infinity. Then one has to perform transformations between the ECI (Earth Centred Inertial system) and the ECEF (Earth Centred Earth Fixed system). The orbital parameters of the satellite constellation are expressed in the ECEF. To realize the ECEF a network of ground stations receiving the GNNS signals has been installed. The GPS uses the World Geodetic System 1984 (WGS-84) and Galileo will use the Galileo Terrestrial Reference Frame (GTRF)~\citep{altamimi09}. These global reference frames are fixed to the Earth (via ground stations) so their precision and stability in time are limited by our knowledge of the Earth dynamics. The main effects are plate tectonic motions, tidal effects on the Earth's crust and variations of the Earth rotation rate. In the WGS-84 the best accuracy achieved is 30~cm~\citep{wgs84}, with an average stability of 4~cm/year~\citep{altamimi09}. The use of other space geodetic techniques - VLBI, SLR and DORIS - is necessary to achieve a high precision ECEF. The International Terrestrial Reference Frame (ITRF), maintained by the International Earth Rotation and Reference Systems Service, combine efficiently these four techniques to reach a stability of about 1~mm/year, inferior by a factor of 10 to science requirements in geology~\citep{altamimi09}.

These considerations led Bartolomé \citet{coll03a} to propose the project ``Syst{\`e}me de Positionnement Relativiste'' (SYPOR), i.e. Relativistic Positioning System, an alternative to the scheme of usual positioning systems. The idea is to give to the constellation of satellites the possibility to constitute by itself a \emph{primary} and \emph{autonomous} positioning system, without any \emph{a priori} realization of a terrestrial reference frame. 
\begin{figure}[ht]
   \begin{minipage}[c]{.46\linewidth}
\includegraphics[width=1.\linewidth]{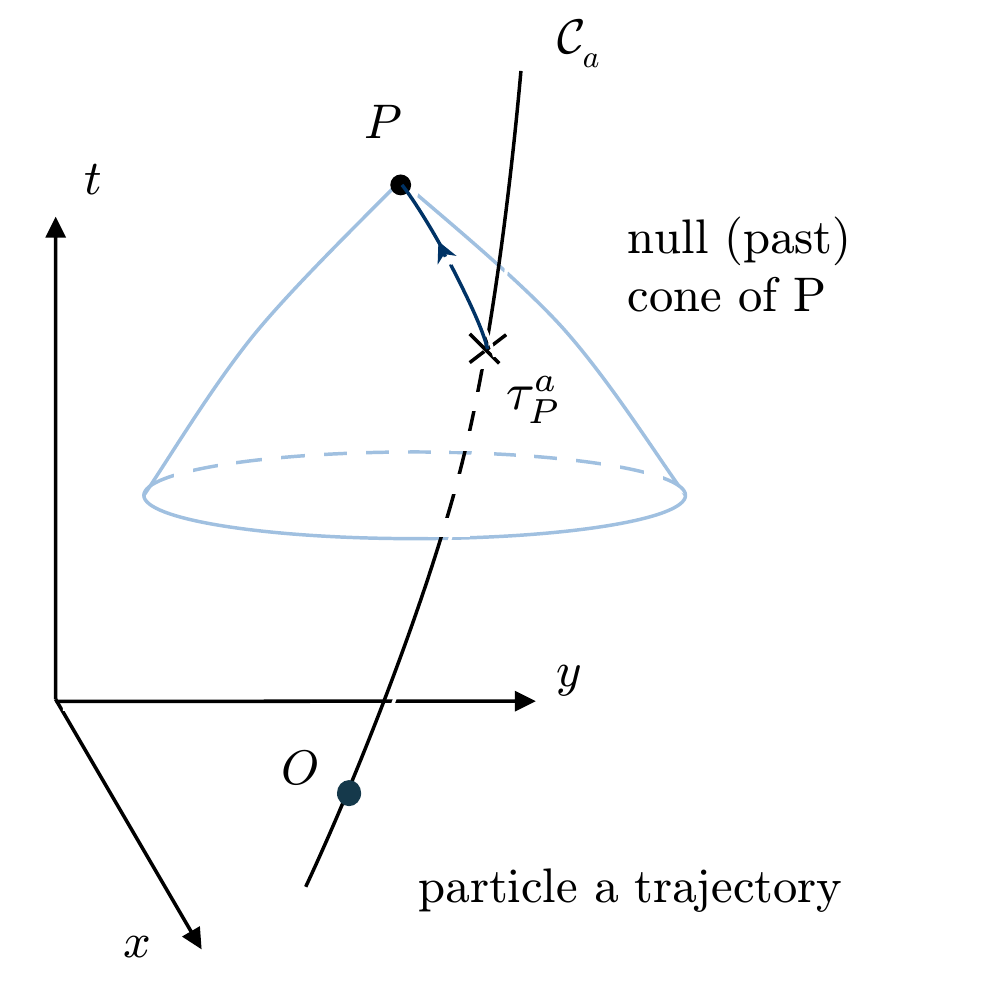}
\caption[Defining null coordinates with the null past cone of a space-time event.]{\label{fig:intro:PC} \footnotesize Defining null coordinates with the null past cone of a space-time event: let $P$ be an event in space-time. $\C_a$ is the worldline of a test particle $a$ parameterized by its proper time $\tau^a$; its origin $O$ is in $\tau^a = 0$. The past null cone of the event $P$ cross $\C_a$ at the proper time $\tau^a_P$. With four different particles with the worldlines $\C_a$ ($a=1,2,3,4$), the past null cone of $P$ crosses the four worldlines in $\tau^1_P$, $\tau^2_P$, $\tau^3_P$ and $\tau^4_P$. Then $(\tau^1,\tau^2,\tau^3,\tau^4)_P$ are the null coordinates of the event $P$.}
   \end{minipage} \hfill
   \begin{minipage}[c]{.46\linewidth}
\includegraphics[width=1.\linewidth]{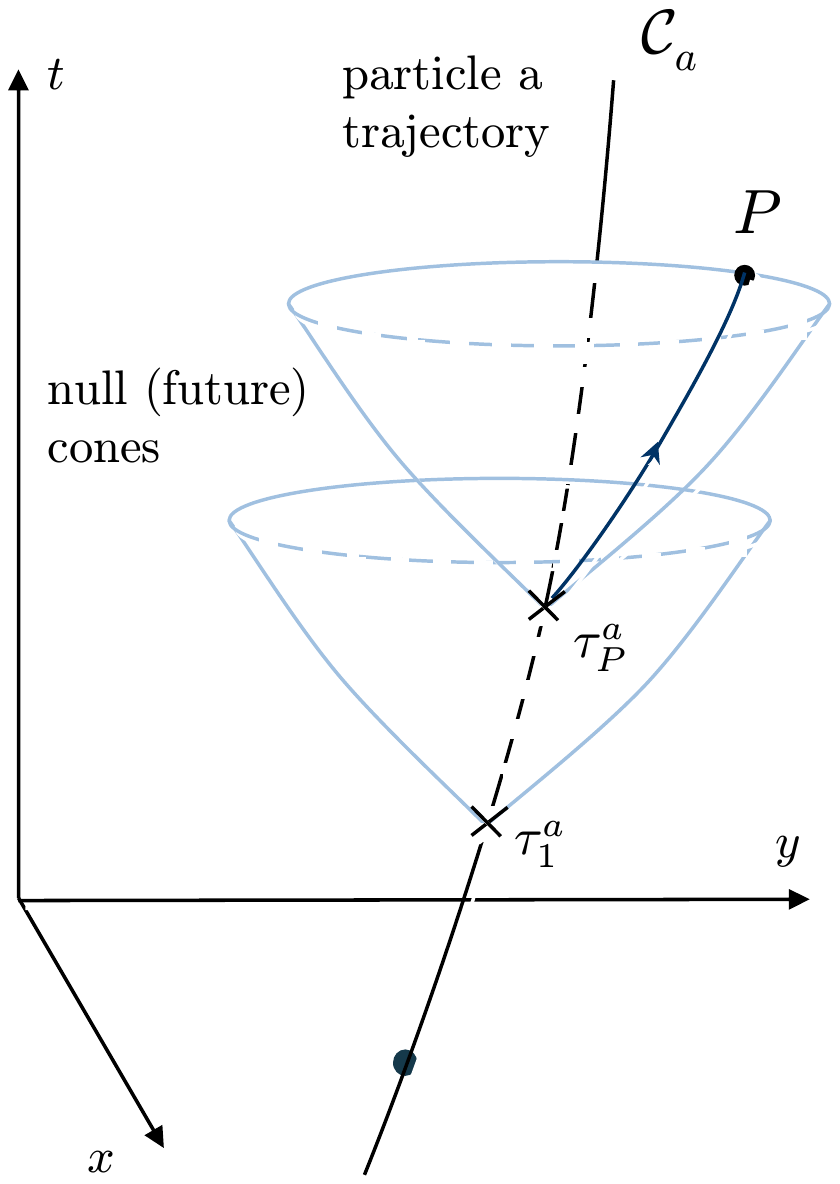}
\caption[Defining null coordinates with four one-parameter families of null future cones.]{\label{fig:intro:FC} \footnotesize Defining null coordinates with four one-parameter families of null future cones: the worldline $\C_a$ defines a one-parameter family of null future cone. These null cones are hypersurfaces of $\tau^a=\text{constant}$. The intersection of four null cones defines the event with coordinates $(\tau^1,\tau^2,\tau^3,\tau^4)$. The four particles can be chosen as four satellites of a constellation of satellites, broadcasting their proper time.}
   \end{minipage}
\end{figure}
The relativistic positioning system is defined with the introduction of \emph{null coordinates}. They have been reintroduced recently by several articles \citep{coll91, rovelli02, blagojevic02, Lachieze-Rey2006}. They have different names in the literature: ``null coordinates'', ``emission coordinates'', ``GPS coordinates'', ``GNSS coordinates''. In this article we use the first name which is a reference to their geometrical properties (see \citet{coll06a} for a detailed article).

The definition of these coordinates is rather simple, but they are a very powerful tool in general relativity. Let us define four particles $a=1,2,3,4$ coupled to general relativity. Their worldlines $\C_a$ are parameterized by their proper time $\tau^ a$. We choose a random origin for $\tau^a=0$ on each worldline. Let $P$ be an arbitrary event. Then the past null cone of $P$ crosses each of the four worldlines in $\tau^a_P$ (see Fig.\ref{fig:intro:PC}). The quadruplet $(\tau^1,\tau^2,\tau^3,\tau^4)_P$ constitutes the complete set of null coordinates of the event $P$. The protocol to define null coordinates can be seen in a different way. The worldline $\C_a$ of the particle $a$ defines a one-parameter family of future null cones, which can be parameterized by proper time $\tau_A$ (see Fig.\ref{fig:intro:FC}). The intersection of four future null cones $\tau^a$ from four worldlines $\C_a$ defines an event with coordinates $(\tau^1,\tau^2,\tau^3,\tau^4)$. A user receiving these signals knows its position in this particular coordinate system.

Coll and collaborators~\citep{coll06b,coll06c} studied relativistic positioning systems in the case of a two-dimensional space-time for geodesic emitters in a Minkowski plane and for static emitters in the Schwarzschild plane. A relativistic positioning system has been studied in the vicinity of the Earth: calculations were performed to first order in a Schwarzschild space-time~\citep{bahder01, ruggiero08}. \citet{bini08} described null coordinates in the local Fermi frame of a user. A ``galactic reference system'' has been studied, where timing signals received by four pulsars were considered as null coordinates~\citep{coll03b,tartaglia10}. The next generation of GNSS could have cross-link capabilities~\citep{dnav09}. Each satellite will broadcast proper time to the other satellites in view, as well as their proper time. With this information, one could \emph{in principle} map the space-time geometry in the vicinity of the constellation of satellites by solving an inverse problem~\citep{tarantola09}.

In this article, we assess the practical feasibility of a relativistic positioning scheme for a constellation of satellites in orbit around the Earth. The Schwarzschild geometry is chosen as a first approximation of the space-time geometry around the Earth. In section~\ref{s:geo} we give solutions of geodesic equations for closed time-like orbits - satellites orbits - and for scattering light-like orbits - photons orbits. In section~\ref{s:simu} we describe two algorithms: (i) how to deduce null coordinates of an event in space-time knowing its Schwarzschild coordinates and orbital parameters of a constellation of four satellites and (ii) how to calculate Schwarzschild coordinates of an event knowing four proper times sent by four satellites, and their orbital parameters (reverse algorithm). We study the speed and accuracy of these two algorithms implemented in Fortran~95. Finally, in section~\ref{s:pert} the effects of non-gravitational perturbations on the relativistic positioning system are assessed.
%
%
\section{Geodesics in the Schwarzschild space-time}
\label{s:geo}
The Schwarzschild space-time is a good approximation of the geometry around the Earth. The local inertial coordinate system is tied to the center of the Earth and oriented into 4 mutually orthogonal directions $t$, $X$, $Y$, and $Z$\footnote{The Earth's local inertial coordinate frame is precessing with respect to the global inertial frame tied to distant stars. The precessions are due to different gravitational perturbations of Earth's multipole moments, gravitational perturbations of the Moon, Sun and planets. These motions can be well modelled, but are not the matter of this article.}. The Schwarzschild space-time is usually represented by the metric in spherical coordinates $t$, $r$, $\theta$, and $\phi$, such that $X=r \sin\theta\cos\phi$, $Y=r \sin\theta\sin\phi$, and $Z=r \cos\theta$ . In these coordinates the metric is
\be \dd s^2 = - C(r) \dd t^2 + C^{-1}(r) \dd r^2 + r^2 \dd \Omega^2 \ee
%
where
\be C(r) = \left( 1-\dfrac{2M}{r} \right) , \ee
$\dd \Omega^2 = \left( \dd \theta^2 + \sin^2 (\theta) \dd \phi^2 \right)$ and $M$ is the Earth mass. We use natural units\footnote{To recover usual units one can replace $M=Gm_{\oplus}/c^2$ when measuring distance and $M=Gm_{\oplus}/c^3$ when measuring time, where $m_{\oplus}$ is the mass of the Earth in the usual units.} $c=G=1$. Geodesics are governed by the Hamiltonian:
\begin{equation}
 H = \frac{1}{2}
   \left[
     -C^{-1}(r) p_t^2 + C(r) p_r^2 +
      \frac{1}{r^2} \left( p_\theta^2 + \frac{1}{\sin^2\theta} \ p_\phi^2 \right)
   \right] ,
\end{equation}
which admits 8 constants of motion~\citep{cadez3}: value of Hamiltonian ($H$) and Lagrangian ($L$), controlling the relation between time and distance, energy ($E=p_t$), three components of angular momentum ($\vec{l}$), longitude of periapsis ($\omega$) and time of periapsis passage ($t_p$).

\begin{figure}[t]
\centering
\includegraphics[scale=0.7]{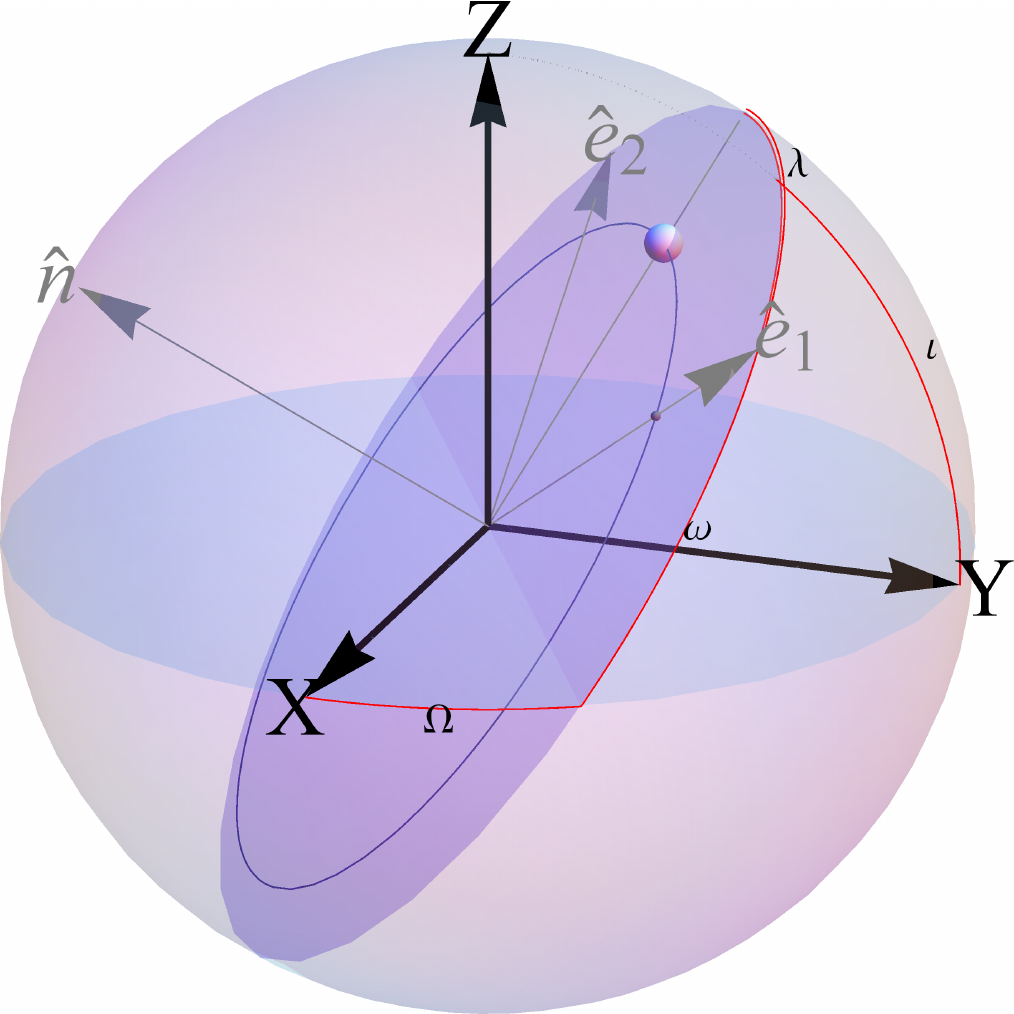}
\caption{The orbital plane in equatorial coordinates: $\vec{n}$ unit normal, $\iota$ inclination, $\Omega$ longitude of the ascending node, $\omega$ longitude of periapsis and $\lambda$ true anomaly.}
\label{f:figl}
\end{figure}

It is convenient to introduce another local inertial (right-handed) orthonormal tetrad $\hat n$, ${\hat e}_1$ and ${\hat e}_2$~(c.f. \reff{f:figl}). $\hat n$ is the constant unit vector pointing in the direction of the angular momentum: $\vec{l}=l \hat{n}$. The two unit vectors ${\hat e}_1$ and ${\hat e}_2$ in the orbital plane are such that ${\hat e}_1$ points in the direction of the initial perigee. The components of these vectors with respect to the local Cartesian coordinate basis are expressed as:
\begin{align}
{\hat e}_1 &=\left (\cos\omega \cos\Omega-\cos\iota\sin \omega\sin\Omega,\ \ \cos\omega\sin\Omega+\cos \iota \sin \omega\cos\Omega ,\sin\iota\sin\omega\right )\nonumber\\
{\hat e}_2 &=\left (\sin\omega \cos\Omega-\cos\iota\cos \omega\sin\Omega,-\sin\omega\sin\Omega+\cos \iota \cos \omega\cos\Omega ,\sin\iota\cos\omega\right )\nonumber\\
\hat n &= \left (\sin\iota\sin\Omega,-\sin\iota\cos\Omega, \cos\iota\right ) ,
\label{eq:vrt_kol_komponente}
\end{align}%
where $\Omega$ is the longitude of the ascending node and $\iota$ is the inclination of the orbit with respect to the $X-Y$ plane.

The only parameter changing along the orbit is the true anomaly $\lambda$. It obeys the differential orbit equation~\citep{cadez3}:
\be
  \frac{\dd u}{\dd\lambda} = \pm\sqrt{A^2-u^2(1-u)+B(1-u)} ,
\label{orbita}
\ee
where $u=2 M/r$, and $A=2ME/l$ and $B=2H(2M/l)^2$ are two constants of motion related to orbital energy and orbital angular momentum. After \refe{orbita} is solved for $u$ as a function of $\lambda$, the orbit can be described with
\beq
\vec{r}(\lambda) = \frac{2M}{u}(\hat{e}_1 \cos \lambda + \hat{e}_2 \sin \lambda)  .
\eq

The spherical coordinates $\theta $ and $\phi$ along the orbit are expressed as\footnote{Note that $\sin\theta=+\sqrt{1-\cos^2\theta}$.}~\citep[Eqs.\hspace{2pt}6~--~16]{cadez3}:
\begin{align}
  \cos\theta & = \sin\iota \sin (\lambda+\omega)\label{eq:incl}\\
  \tan \frac{\varphi -\Omega}{2} & = \frac{\cos\iota \sin(\lambda + \omega)}
  					  {\sin\theta + \cos(\lambda + \omega)}
  					  \label{eq:ascnode}
\end{align}

Time and proper time obey the following differential equations:
\begin{subequations}
\begin{align}
	\frac{\dd t}{\dd u} & = \frac{2MA}{	u^2 (1-u)\sqrt{A^2 - u^2(1-u) + B(1-u)}	}\label{eq:times1}\\
	\frac{\dd \tau}{\dd u} & = \frac{2MA}{E}	\frac{1}{	u^2\sqrt{A^2 - u^2(1-u) + B(1-u)}	}  .
\label{eq:times}
\end{align}
\end{subequations}

The differential equations~\eqref{orbita} and~\eqref{eq:times1} are formally the same for light-like (where $B=0$) and time-like orbits. However, solutions depend on the type of orbit, e.g. closed, scattering or plunging. In the following subsections we give solutions for closed time-like orbits and scattering light-like orbits, which can be used to model satellites and photons trajectory in a GNSS.

\subsection{Time-like geodesics}
GNSS satellites are on closed time-like orbits. After solving \refe{orbita} for such case, we obtain the orbit equation
\beq
	u(\lambda) = U_2-(U_2-U_3)\cn^2
		  \left(
		  \elK(m_a) + \frac{\lambda}{2 n_a} \bigg| m_a
	\right)\ , \label{eq:u_lambda}
\eq
where $U_1$, $U_2$ and $U_3$ are the roots of the polynomial $P(u)=A^2-u^2(1-u)+B(1-u)$, and $n_a$ and $m_a$ are functions of them. $\elK$ and $\cn$ are the complete elliptic integral of the first kind and Jacobian elliptic function respectively. The true anomaly $\lambda$ is defined as in the Keplerian case; at periapsis it has values $\lambda_p = 4n_a \elK (m_a)k$, where $k$ is an integer. $U_2$ and $U_3$ are related to the radii of the periapsis $\rp = 2M/U_2$ and apoapsis $\ra = 2M/U_3$ respectively. For circular orbits $U_2 = U_3$, and the orbit equation reduces to $r = const. = \ra = \rp$.

Schwarzschild time and proper time along the orbit are obtained by integrating equations \refee{eq:times1}{eq:times}. Their expressions can be found in~\citet{cadez4}.

\begin{figure}[t]
\includegraphics[height=0.3\textwidth]{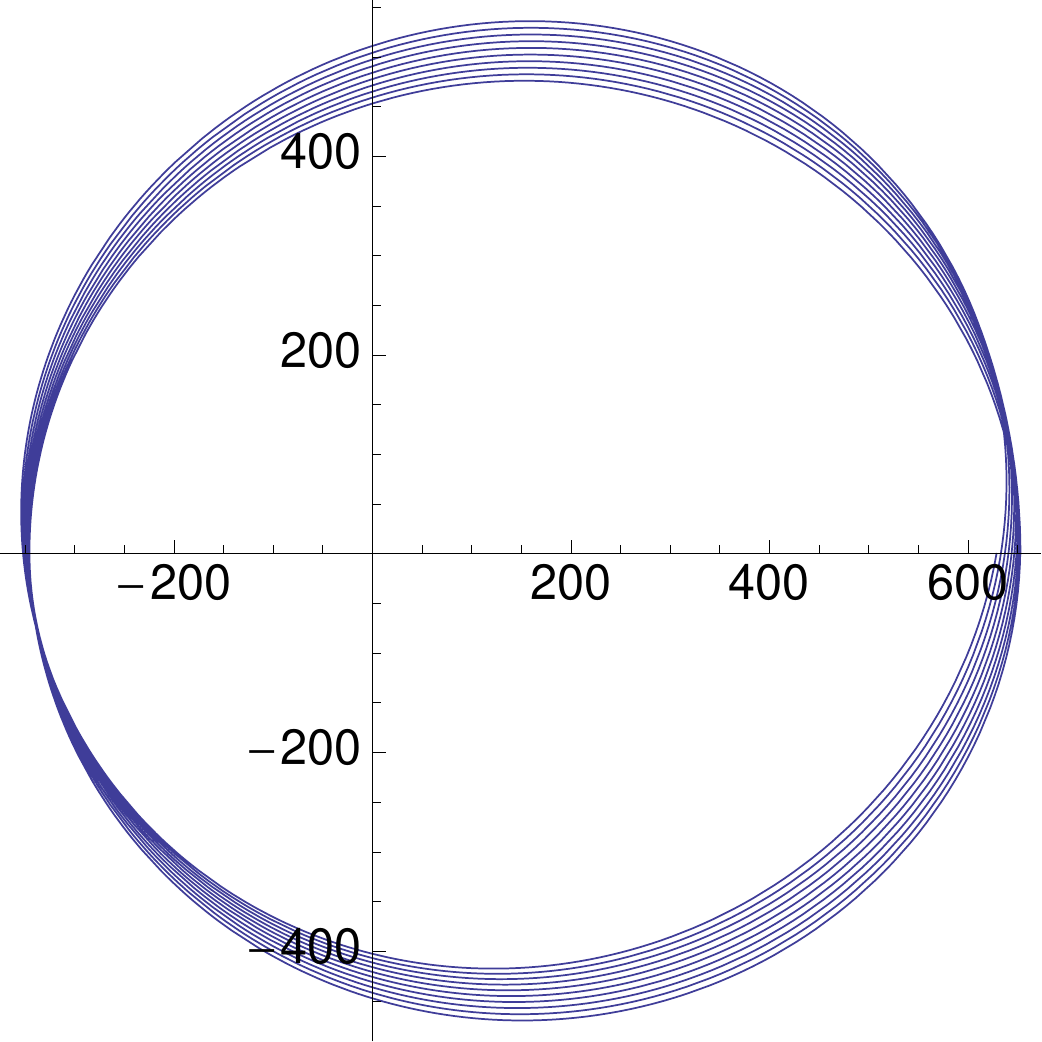}
\hspace{0.5cm}
\includegraphics[width=0.3\textwidth]{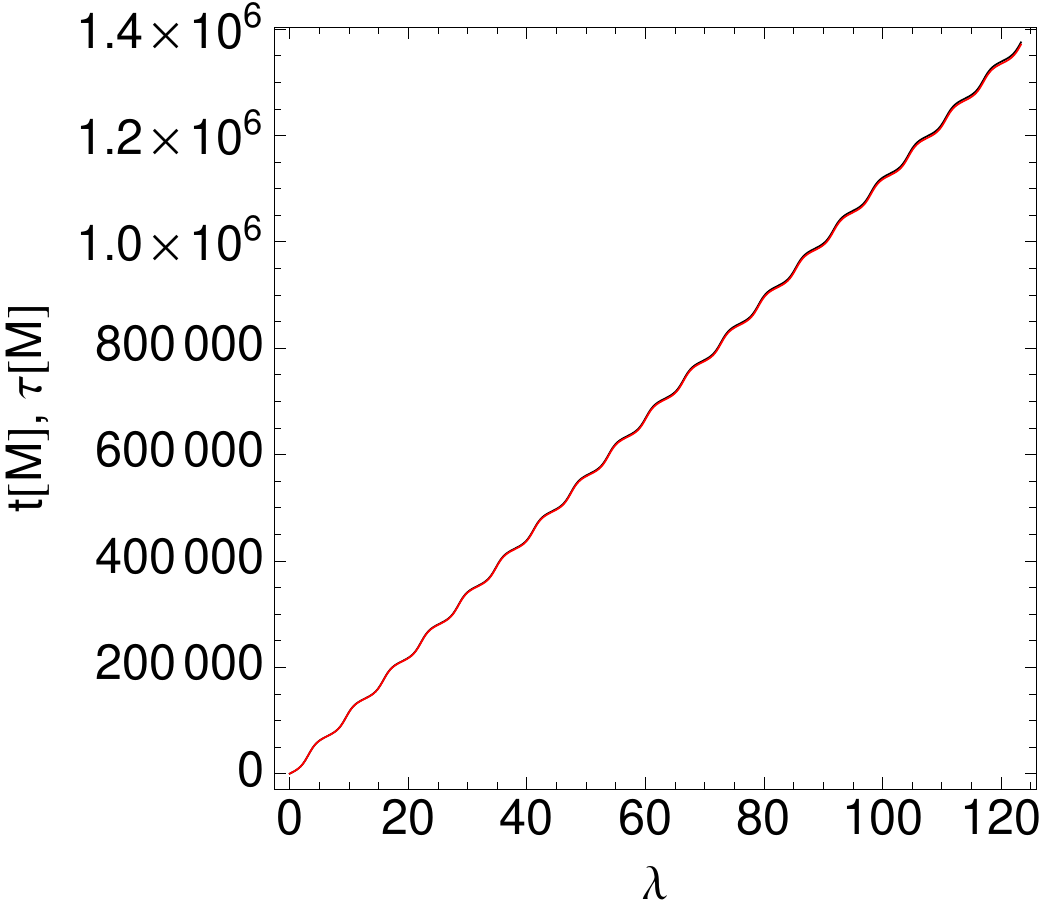}
\hspace{0.5cm}
\includegraphics[width=0.3\textwidth]{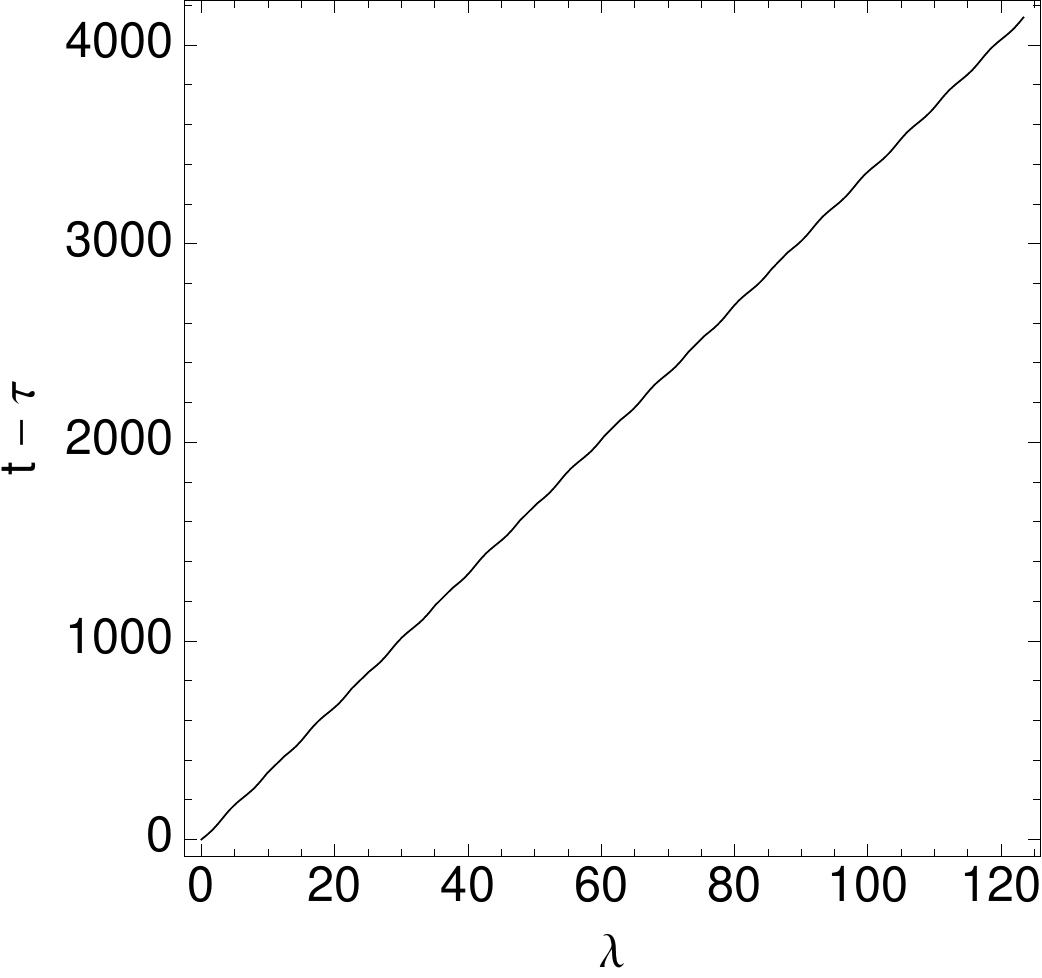}
\caption{Left: A time-like orbit with semi-major axis $a=500\ \rg$ and eccentricity $\eksc=0.3$. Middle: Time (black) and proper time (red) for the same orbit. Right: The difference between time and proper time for the same orbit. All values of $r$, $t$, and $\tau$ are in units of $M$.}
\label{fig:timelike_example}
\end{figure}
A solution of the time-like geodesic equation is illustrated in \reff{fig:timelike_example}. Orbital parameters were intentionally chosen such that relativistic effects are clearly visible, i.e. periapsis precession and the difference between time and proper time. For the values of orbital parameters of the Galileo satellites, the difference between coordinate time and proper time goes up to $10\ {\mu s}$ per orbit.


%

\subsection{Light-like geodesics and ray-tracing}
\label{sec:light}
In the eikonal approximation, electromagnetic signals sent by satellites follow null scattering geodesics. After solving \refe{orbita} with $B=0$ for such cases, we obtain the orbit equation
\beq
	u(\lambda) = u_2-(u_2-u_3)\cn^2
		  \left(
		  \elK(m) + \frac{\lambda}{n} \bigg| m
	\right)\ , \label{eq:u_lambda_light}
\eq
where the true anomaly $\lambda$ takes values on the interval $\lambda \in (\elF(\chi_{max}|m) - \elK(m),\elF(\chi_{min}|m) - \elK(m))$. Here $\chi_{min} = \arccos \left(\sqrt{u_2/(u_2 - u_3)}\right)$, $\chi_{max} = \arccos \left(-\sqrt{u_2/(u_2 - u_3)}\right)$, $\elF$ is the elliptic integral of the first kind, and $u_1$, $u_2$, $u_3$ are the roots of the polynomial $P(u)=A^2-u^2(1-u)$. The constants $u_1$, $u_2$, $u_3$, $m$, $n$ depend only on one constant of motion - $A=2ME/l$, which is the inverse of the impact parameter. 

Coordinate time and proper time as a function of $\lambda$ is obtained by integrating equations \refe{eq:times1} and \refe{eq:times} \citep{cadez4}. In the case of GNSS, the gravitational field is very weak so photon orbits are essentially straight lines. Differences between the relativistic and non-relativistic time-of-flight are of the order of 1~ns when considering a signal travelling from a satellite to the user.

\begin{figure}[t]
\centering
\includegraphics[width=0.5\textwidth]{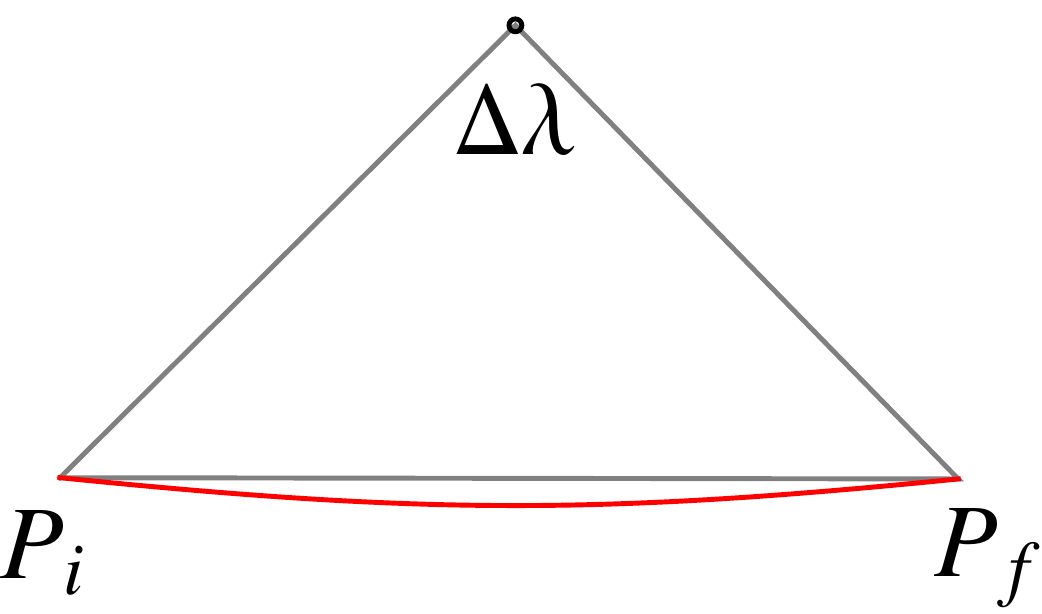}
\caption{Determining a light-like geodesic between points $\pp_i$ and $\pp_f$ from known $\Delta\lambda$, $r_i$, and $r_f$. The calculated orbit is shown in red. This case corresponds to a strong gravitational field in order to see the effect.}
\label{fig:rt}
\end{figure}
Calculating null coordinates of an event in space-time requires one to be able to follow light rays emitted from the satellites towards that event. Thus, one would like to find the quickest way to determine all the constants of motion of a null geodesic that connect a given initial point $\pp_i=(t_i,r_i,\theta_i,\varphi_i)$ and a final point $\pp_f=(t_f,r_f,\theta_f,\varphi_f)$. \citet{2005PhRvD..72j4024C} have proposed a ray-tracing method which completely solves this problem in the Schwarzschild space-time. The method uses spherical trigonometry and the analytic expression of the orbit equation \refe{eq:u_lambda_light}. An example of orbit determination from $\Delta\lambda$, $r_i$, and $r_f$ is shown in \reff{fig:rt}.

%
%
%
\section{Simulation of the constellation of satellites}
\label{s:simu}
\subsection{Determining null coordinates}
\label{sec:Schw_null}
\begin{figure}[t]
\centering
\includegraphics[width=0.4\textwidth]{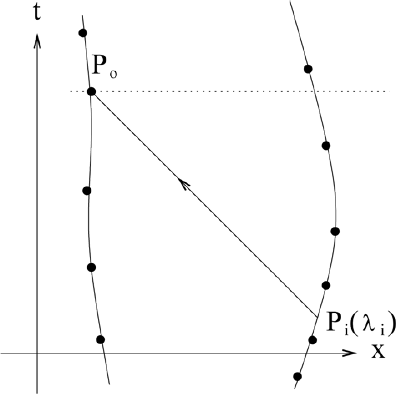}
\caption{Determining null coordinates from Schwarzschild coordinates. The signal is received at $P_o$. The point of emission $P_i$ has to be determined by connecting the two points with a light-like geodesic (diagonal line). For clarity, only one spatial coordinate is shown.}
\label{fig:Schw_null}
\end{figure}
We assume that the constants of motion of all satellites are known. Their trajectory can be calculated and parameterized by their true anomaly $\lambda$. Let us define a user at the event $\pp_o = (t_o,x_o,y_o,z_o)$. He receives signals from four satellites that were sent at $\pp_i = (t_i,x_i,y_i,z_i)$, corresponding to $\lambda_i$ ($i=1,...,4$). Null coordinates of the user at $\pp_o$ are the proper times $\tau_i (\lambda_i)$ of the satellites at $\pp_i$ (see \reff{fig:Schw_null}). We calculate $\lambda_i$ at the emission point $\pp_i$ using the equation that connects $\pp_o$ and $\pp_i$ with a light-like geodesic
\beq
t_o - t_i(\lambda_i) = T_f(\vec{R}_i(\lambda_i), \vec{R}_o)\ ,
\eq
where $\vec{R}_i=(X_i,Y_i,Z_i)$ and $\vec{R}_o = (X_o,Y_o,Z_o)$ are respectively the spatial vectors of the satellites and the user. The function $T_f$ calculates the time-of-flight of photons between $\pp_o$ and $\pp_i$ as described in \citet{2005PhRvD..72j4024C}. This equation can be very efficiently solved with known algorithms, e.g. Newton's method. Once the value of $\lambda_i$ is determined, it is straightforward to calculate the proper time of emission $\tau_i$ for each satellite and thus obtain null coordinates of the user at $\pp_o = (\tau_1,\tau_2,\tau_3,\tau_4)$.

\subsection{Calculating Schwarzschild coordinates}
\label{sec:null_Schw}
Here we solve the inverse problem: calculate Schwarzschild coordinates of the event $\pp_o$ from $(\tau_1,\tau_2,\tau_3,\tau_4)$ sent by the four satellites. As we assume that the constants of motion of all satellites are known, we can deduce their space-time positions $\pp_i=(t_i,x_i,y_i,z_i)$ from proper times $\tau_i$. Events $\pp_i=(t_i,x_i,y_i,z_i)$ and $\pp_o=(t_o,x_o,y_o,z_o)$ are connected with light-like geodesics. In a flat space-time, they solve the four equations
\be
	t_o - t_i = \sqrt{(x_i - x_o)^2 + (y_i - y_o)^2 + ( x_i - x_o)^2} .
\label{eq:distance}
\eq

\begin{figure}[t]
\centering
\includegraphics[height=0.6\textwidth]{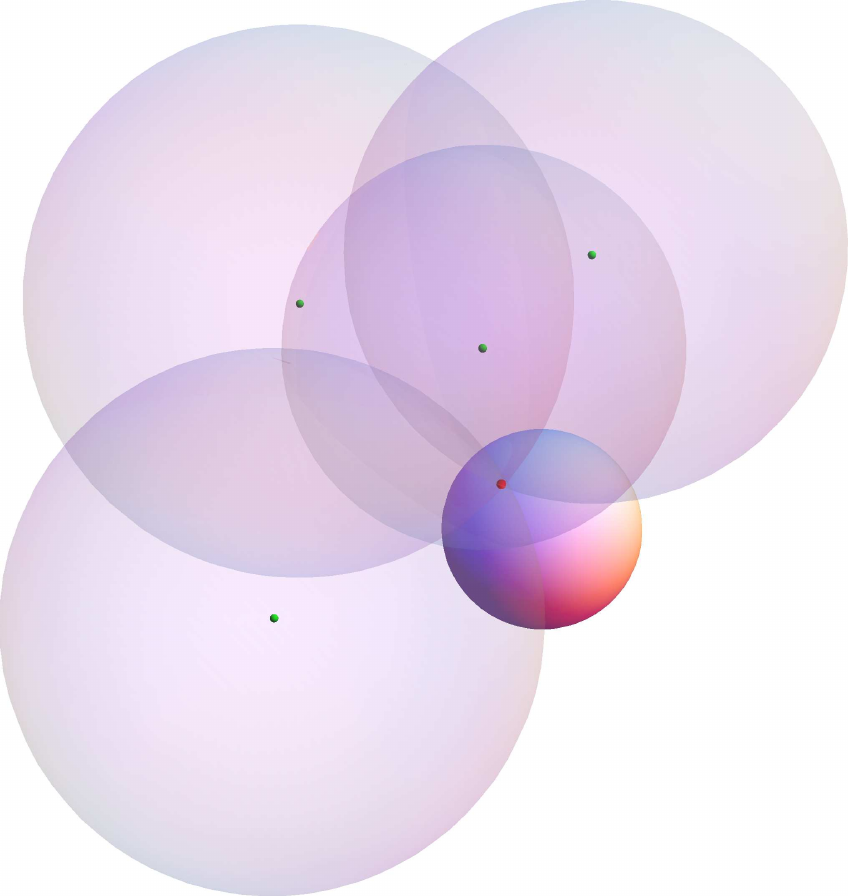}
\caption{The four green points represent the four satellites at $\vec{R}_i$ and the red point is the user.}
\label{fig:4spheres}
\end{figure}

These four equations can be solved for $(t_o,x_o,y_o,z_o)$ by a geometric construction. Let $\vec{R}_i =(X_i,Y_i,Z_i)$ be the spatial coordinates vectors of the satellites at $\pp_i$. The situation is illustrated in \reff{fig:4spheres}, where the four green points represent the four satellites at $\vec{R}_i$ and the red point is the user. 

The 4 spheres centred at $\vec{R}_i$ have radii $(t_o - t_i)$. Thus, the user is at the intersection of the four spheres. To find his position, we proceed as follows: suppose that the user's dead reckoning coordinates are $(t_o^{(0)},x_o^{(0)},y_o^{(0)},z_o^{(0)})$. The radii of the four spheres centred at $\vec{R}_i$ would then be $(t_o^{(0)} - t_i)$. In general, these four spheres have no common point. However, the dead reckoning position can always be chosen in such a way that any two spheres intersect. Consider the planes defined by the circle of intersection of sphere 1 and 2 and sphere 3 and 4. These two planes generally intersect along a straight line; call it line 1. Next consider the intersection of spheres 1 and 3, and spheres 2 and 4. The corresponding planes intersect along a straight line called line 2. If \refe{eq:distance} is satisfied, then line 1 and line 2 intersect at the position of the user. However, since $(t_o^{(0)}, x_o^{(0)},y_o^{(0)},z_o^{(0)})$ is not yet the solution of \refe{eq:distance}, the lines 1 and line 2 generally bypass each other. We calculate the positions on both lines where they meet at closest distance. The geometrical center of the two positions is taken as the better approximation for the spatial position of the user $\vec{R}_o^{(1)}$, and the distance between the two points of closest passage $(\dd\vec{P})$ is a measure of how close we are to the solution. In the next step, we repeat the procedure with a slightly different $t_o$ to find the derivative of $\dd\vec{P}$ with respect to $t_o$. This derivative is then used in Newton's method to find a new $t_o^{(1)}$ with a smaller $\dd\vec{P}$. With this value a new spatial position $\vec{R}_o^{(2)}$ is calculated and the procedure is repeated until $\dd\vec{P}$ becomes less than the value prescribed by the accuracy requirement. After $n$ steps $(n\sim 5)$, the procedure converges to $\{t_o^{(n)}, \vec{R}_o^{(n)}\}$, the solution of \refe{eq:distance}. Note, however, that a unique solution exists only if the normals to all planes whose intersections define lines 1 and 2 do not lie in a plane, i.e. if the four satellites do not lie in one plane.

However, eq.~\refe{eq:distance} does not take into account the gravitational time delay and the bending of light, so our result is not exact. For realistic positions of Galileo satellites the position error is a few $10^{-9}$ orbital radii, i.e. a few centimetres. The final correction to the position is made by solving the relativistic equations of light propagation from the satellites to the user in the following form:
\beq
	t_o - t_i = T_f(\vec{R}_i, \vec{R}_o)\ ,
\label{eq:rel_dist}
\eq
where $T_f(\vec{R}_i, \vec{R}_o)$ calculates the coordinate time for the light to travel from $\vec{R}_i$ to $\vec{R}_o$. Taking $\{t_o^{(n)}, \vec{R}_o^{(n)}\}$ as an initial approximation, we use a generalization of the classical stellar navigation solution to solve \refe{eq:rel_dist}. This equation is written in the form:
\bea
	t_o^{(n)} + \dd t - t_i &=& T_f(\vec{R}_i , \vec{R}_o^{(n)} + \dd \vec{R}) \nonumber \\[.3cm]
	& = &  T_f(\vec{R}_i, \vec{R}_o^{(n)}) +  \vec{\nabla}_o T_f (\vec{R}_i , \vec{R}_o^{(n)}) \cdot \dd \vec{R} + \Ol (\dd \vec{R}^2) ,
\eea
where $\vec{\nabla}_o$ is the gradient with respect to the position of the user $\vec{R}_o$. We now assume that the gravitational field is weak, i.e. $|M/R| \ll 1$. Then 
\be T_f(\vec{R}_i, \vec{R}_o) = | \vec{R}_o - \vec{R}_i | + \Ol(M) \nonumber \ee
and
\bea
\vec{\nabla}_o T_f (\vec{R}_i , \vec{R}_o) &=& \dfrac{\vec{R}_o - \vec{R}_i}{| \vec{R}_o - \vec{R}_i |} + \Ol(M/R) \nonumber \\[.3cm]
&=& \hat{u}_i + \Ol(M/R) ,
\eea
where $\hat{u}_i$ is the unit vector pointing from satellite $i$ to the supposed position of the user. Eq.~\refe{eq:rel_dist} now becomes a set of four linear equations for $\dd t$ and the three components of $\dd \vec{R}$:
\beq
	t_o^{(n)} - t_i  - T_f(\vec{R}_i , \vec{R}_o^{(n)}) = - \dd t + \vec{u}^{(n)}_i \cdot \dd \vec{R} .
\label{eq:rel_final}
\eq
The error of the corrected position decreases by 6 to 9 orders of magnitude -- bringing the Galileo position error to the order of micrometers. If necessary, this error can be decreased even further by solving  \refe{eq:rel_final} again with $t_o^{(n+1)}= t_o^{(n)} + \dd t$ and 
\beq
	\vec{R}_o^{(n+1)} = \vec{R}_o^{(n)} + \dd \vec{R} \ .
\eq

\subsection{Accuracy and speed of algorithms}
\begin{figure}
\centering
\includegraphics[width=0.45\textwidth]{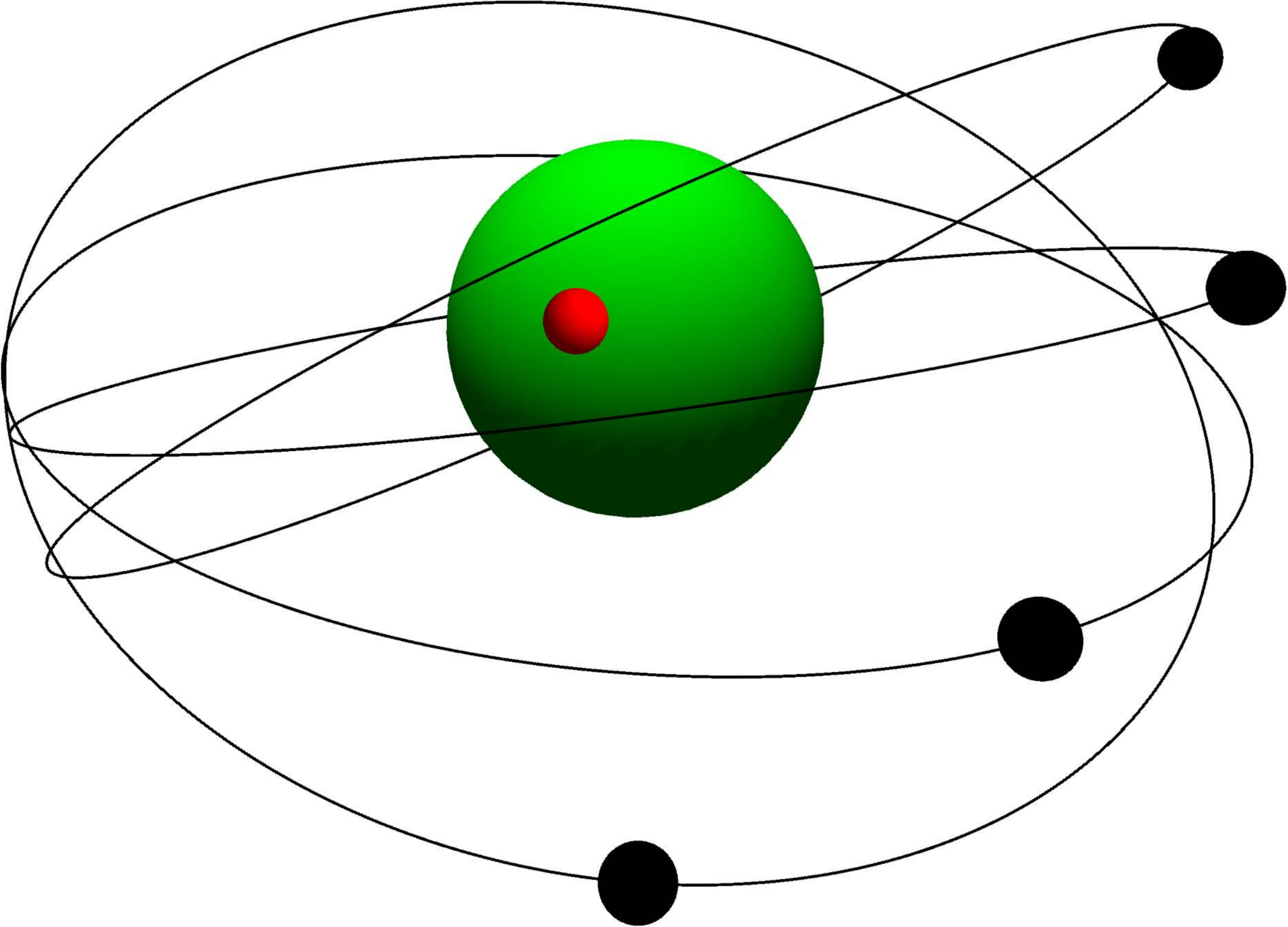}
\caption{The orbits of 4 satellites around the Earth. The satellites at initial positions are marked with black dots, and the user on Earth's surface with a red dot. The green sphere represents the Earth. The sizes of satellites are not to scale.}
\label{fig:satellites}
\end{figure}
\begin{table}[t]
\centering
	\begin{tabular}{c c c c c c c}
	\hline
	\hline
		satellite & $\Omega\ [ \degr]$ & $\omega\ [ \degr]$ & \incl$\ [ \degr]$ & $a\ [\mathrm{\rg}]$ & \eksc & $t_p\ [\mathrm{\rg/c}]$\\
	\hline
		1 & 0 & 0 & 45 & $5\times 10^9$ & $1.1\times 10^{-9}$ & 0\\
		2 & 30 & 5 & 45 & $5\times 10^9$ & $1.1\times 10^{-9}$ & 0\\
		3 & 60 & 10 & 45 & $5\times 10^9$ & $1.1\times 10^{-9}$ & 0\\
		4 & 90 & 15 & 45 & $5\times 10^9$ & $1.1\times 10^{-9}$ & 0\\
	\hline
	\hline
	\end{tabular}
\caption{Orbital parameters for 4 satellites: longitude of ascending node ($\Omega$), longitude of perigee ($\omega$), inclination (\incl), major semi-axis ($a$), eccentricity (\eksc), and time of perigee passage ($t_p$).}
\label{tab:parameters}
\end{table}
The above algorithms were tested in a simulation of 4 satellites $S_i$ ($i=1,...,4$) in orbit around the Earth communicating with a static user on the Earth's surface\footnote{The problem of the atmospheric perturbations is not in the scope of this paper.}. The only input parameters are the orbital parameters of the four satellites (table~\ref{tab:parameters}) and the coordinates of the user. The orbits for the four satellites are illustrated in \reff{fig:satellites}. The Schwarzschild coordinates of the user are
\be r_o=1.595\cdot 10^9\ M \ , \ \theta_o=43.97\degr \ , \ \phi_o=14.5\degr . \nonumber \ee
The simulation runs in the following way. At every time-step
\beq \nonumber
	t^{(n)} = t^{(n-1)} + \Delta t\ ,\hspace{0.5cm}\ n=2,3... ,
\eq
where $\Delta t=6\cdot 10^{12}\ M$ and $n$ counts the time-steps of the simulation:
\begin{enumerate}
\item We calculate null coordinates $(\tau_1^{(n)},\tau_2^{(n)},\tau_3^{(n)},\tau_4^{(n)})$ of the user from his Schwarzschild coordinates $(t^{(n)},x_o^{(n-1)},y_o^{(n-1)},z_o^{(n-1)})$, as described in \refs{sec:Schw_null}. (For $n=1$ we choose $(t^{(1)},x_o^{(0)},y_o^{(0)},z_o^{(0)})=(0,r_o,\theta_o,\phi_o)$).
\item From previously calculated null coordinates $(\tau_1^{(n)},\tau_2^{(n)},\tau_3^{(n)},\tau_4^{(n)})$, we calculate Schwarzschild coordinates $(t_i^{(n)},x_i^{(n)},y_i^{(n)},z_i^{(n)})$ for every satellite $S_i$ at its new position $\tau^{(n)}_i$ by numerically solving the equation $\tau(\lambda^{(n)}_i) = \tau^{(n)}_i$ for $\lambda^{(n)}_i$.
\item From these Schwarzschild coordinates $(t_i^{(n)},x_i^{(n)},y_i^{(n)},z_i^{(n)})$, we calculate Schwarzschild coordinates $(t_o^{(n)},x_o^{(n)},y_o^{(n)},z_o^{(n)})$ of the user as described in \refs{sec:null_Schw}.
\end{enumerate}

The numerical code of this simulation is written in Fortran 90 and is publicly available on the website \url{atlas.estec.esa.int/ariadnet}. This code can be easily generalized to include a moving user, more satellites or communications between all the satellites.


\paragraph{Accuracy} The accuracy of these algorithms is tested by comparing the initial Schwarzschild coordinates of the user with his Schwarzschild coordinates calculated at each time step. As the user is static, his coordinates $(x_o,y_o,z_o)$ should be constant during the simulation.

In table \ref{tab:rel_errors} we show the relative errors of all the coordinates, defined as
\be \label{e:errtime}
	\epsilon_t^{(n)}  = \frac{ t^{(n)} - t_o^{(n)}}{t^{(n)}}
\ee
and
\be \label{e:errpos}
	\epsilon_x^{(n)}  = \frac{ x_o^{(1)} - t_o^{(n)}}{x_o^{(1)}} , \
	\epsilon_y^{(n)}  = \frac{ y_o^{(1)} - t_o^{(n)}}{y_o^{(1)}} , \
	\epsilon_z^{(n)}  = \frac{ z_o^{(1)} - t_o^{(n)}}{z_o^{(1)}} .
\ee
\begin{table}[t]
\centering
	\begin{tabular}{c c c c c}
	\hline
	\hline
		$n$ & $\epsilon_t$ & $\epsilon_x$ & $\epsilon_y$ & $\epsilon_z$\\
	\hline
430 &    1.93758E-32 &   -6.15473E-27 &   -7.34498E-26 &   -4.86318E-26\\
431 &   -2.99206E-32 &   -1.78472E-26 &   -7.64415E-26 &    1.84489E-25\\
432 &   -1.46740E-31 &   -4.82243E-26 &   -1.13881E-25 &    7.31360E-25\\
433 &   -1.72335E-32 &   -1.09471E-26 &   -7.03668E-26 &    1.27280E-25\\
434 &   -1.01827E-32 &   -7.65914E-27 &   -5.69550E-26 &    9.27816E-26\\
	\hline
	\hline
	\end{tabular}
\caption{Relative errors of the coordinates as defined in \refee{e:errtime}{e:errpos}, for the last five time-steps of the simulation, which is equivalent to more than one orbit of the satellites.}
\label{tab:rel_errors}
\end{table}
The relative error of the coordinate $t$ is $\sim 10^{-32}$, and relative errors of the coordinates $x$, $y$, and $z$ are few orders of magnitude larger, i.e. $\sim 10^{-25} - 10^{-27}$.

The algorithm for determining Schwarzschild coordinates from null coordinates (\refs{sec:null_Schw}) works only if the four satellites are not in the same plane. This may happen during a simulation in which case the resulting position has a very large error or is left undetermined. As an example, we show in table \ref{tab:rel_errors2} relative errors where the positions cannot be determined for a given configuration of the satellites, which is reflected in a jump in the relative errors from $\sim 10^{-31}$ to $\sim 10^{-2}$ for the coordinate $t$, and from $\sim 10^{-25}$ to $\sim 10^{+3}$ for the coordinate $z$. In the case of GNSS satellites such a situation should never occur, since there are always more than four satellites visible and it should always be possible to choose four that do not lie in the same plane.
\begin{table}[h]
\centering
	\begin{tabular}{c c c c c}
	\hline
	\hline
		$n$ & $\epsilon_t$ & $\epsilon_x$ & $\epsilon_y$ & $\epsilon_z$ \\
	\hline
  7 &   -1.11018E-30 &    3.21695E-27 &   -8.46151E-26 &    1.88565E-25\\
  8 &   -9.83688E-31 &    3.86661E-27 &   -9.02631E-26 &    2.02826E-25\\
  9 &   -9.29760E-31 &    3.02395E-27 &   -8.94093E-26 &    2.02005E-25\\
 10 &   -5.15060E-02 &   -7.85972E+02 &   -7.40892E+01 &   -2.31309E+03\\
 11 &   -1.51029E-02 &   -2.56711E+02 &   -2.44827E+01 &   -7.56241E+02\\
 12 &   -9.12949E-03 &   -1.71063E+02 &   -1.66154E+01 &   -5.04159E+02\\
	\hline
	\hline
	\end{tabular}
\caption{Relative errors for a planar configuration of satellites.}
\label{tab:rel_errors2}
\end{table}


\paragraph{Speed of calculations} The simulation was tested on a PC with an Intel Core2 Quad CPU Processor Q6600 at 2.4 GHz and 4GB RAM. The OS was Linux x86\_64 with kernel 2.6.28-18-generic, with the Intel Fortran Compiler 10.1. The maximum number of time-steps is $n_{max}=434$, equivalent to more than one orbit of the satellites. The calculation for steps 1 to 3 described above repeated 434 times (i.e. for 434 time-steps) takes 67.6 seconds to execute (0.1558 seconds/time-step). The time of calculation for steps 2 and 3 (repeated 434 times) is 26.6 seconds (0.0682 seconds/time-step). These two steps are actually the ones that the users will perform to determine their positions.

The simulation was also tested on a laptop with an Intel Core2 Duo CPU Processor P8600 at 2.4 GHz and 4GB RAM. The OS is the same as before, but with the Intel Fortran Compiler 11.1. In this case, some processor-specific compiler optimizations are enabled. The maximum number of time-steps is the same as before: $n_{max}=434$. The time of calculation for steps 1-3 described above repeated 434 times (i.e. for 434 time-steps) is 61.9 seconds (0.1426 seconds/time-step). The time of calculation for steps 2 and 3 (repeated 434 times) is 26.5 seconds (0.0588 seconds/time-step).

In both cases, a 25 -- 32-digit accuracy is achieved when determining the space-time position of the user.

%
%
\section{Effects of non-gravitational perturbations}
\label{s:pert}
The important non-gravitational perturbations are those governed by stochastic noise, i.e. by phenomena that cannot be predicted in advance. Clock noise, solar radiation pressure, solar wind pressure and collisions with interplanetary dust particles are sources of such noise. The nature of these forces and also the nature of the uncertainty in knowing their magnitude and direction are quite different. 

\paragraph{Clock errors} Clock errors affect the position determination by giving false information on the time of flight from the satellite to the user. Assuming that a typical clock error can be described as a flicker noise with a standard deviation error of 1~ns/day, we can assign a displacement error of $d_{\mathrm t}=$0.3~m per day to stochastic clock perturbations.

\paragraph{Solar radiation} The solar radiation pressure produces a force
\be \vec F_{rp}=\left(1+({\underline{\underline \eta}}-1)a\right )\cdot\dfrac{L_\odot}{4\pi r^3 c}\vec r {\cal A} , \ee
where $\vec r$ is the radius vector from the Sun to the satellite, $\cal A$ is the cross section of the satellite from the impinging direction, $a$ is the effective albedo of the satellite (generally depending on the angle of incidence of the solar light on the satellite) and ${\underline{\underline \eta}}$ is a tensor attached to the satellite, measuring the effective momentum of reflected light. Assuming that $a$, ${\underline{\underline \eta}}$ and $\cal A$ are constants, the effect can be calculated very precisely since the solar luminosity is a well known constant. In this case the main effect of solar radiation pressure is a precession of the satellite orbital angular momentum with an angular velocity
\be {\vec \Omega}_{rp}= \dfrac{\Omega_{\mathrm{year}}}{\omega_s^2 r_s} \cdot \frac{\hat n \times {\vec F}_{rp}}{m} , \ee
\begin{figure}[t]
\centering
\includegraphics[width=0.65\textwidth]{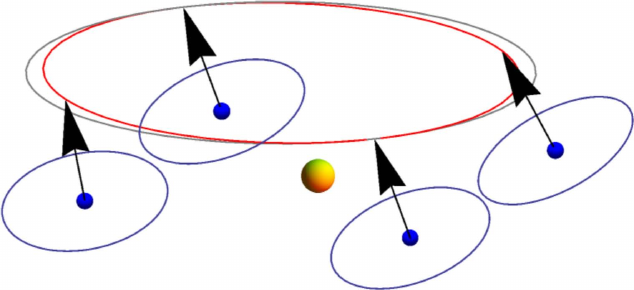}
\caption{As the Earth (blue sphere) moves about the Sun (yellow sphere), the solar radiation pressure makes the satellites' orbital angular momentum precess so that the tip of the orbital angular momentum follows the red ellipse; the gray circle shows where the tip of angular momentum would be without precession.}
\label{fig:RadPressure}
\end{figure}
where $\hat n$ is defined in section~\ref{s:geo}, m is the mass of the satellite, $\Omega_{\mathrm{year}}$ is the angular velocity of the solar radiation force with respect to the orbital angular velocity of a satellite, i.e. $\Omega_{\mathrm{year}} \simeq \frac {2\pi}{1 \ \mathrm{year}}$, $\omega_s= 2\pi \ P_s$ (where $P_s \sim \frac{1}{2} \ \mathrm{day}$) is the orbital angular velocity of the satellite and $r_s$ is the distance of the satellite from the center of the Earth. This effect, schematically shown in Fig.~\ref{fig:RadPressure}, makes the orbital angular momentum oscillate with the orbital period of the Earth around the Sun. The displacement of the satellite's position due to solar radiation pressure as a function of time can be written as
\be d_{rp}=\frac{F_{rp}}{4\pi^2 m}P_s^2\sin \left (\Omega_{\mathrm year}t\right )\cos(\omega_s t+\delta) . \ee
The amplitude of this term can be considered as a measure of the solar radiation strength. If we assume the satellite to have a mass of 600~kg, a
cross sectional area of 0.5~m$^2$ and take ${\cal F}_\odot=1300$~W/m$^2$ for the solar constant, we obtain $d_{rp}\sim 0.17$~m.

A realistic calculation of the effect of the solar radiation pressure is somewhat more complicated, since $a$, ${\underline{\underline \eta}}$ and $\cal A$ depend on the orientation of the satellite with respect to the direction to the Sun. It is probably safe to assume that this dependence can be known  to better than 1\% or maybe even 0.1\% in which case the unpredictable stochastic part of the effect of solar radiation pressure would be always below millimeters.

\paragraph{Solar wind} The perturbing force due to solar wind pressure is similar to the radiation pressure force since it originates from the same direction. However the pressure term $\frac{L_\odot}{4\pi r^3c}\vec r$ must be replaced by $\frac{\dot M }{4 \pi r^2}\vec v$, where $\dot M$ and $\vec v$ are respectively the solar wind mass loss rate and velocity. Their typical values found in literature are: $\dot M\sim 4\times 10^{19}\ \mathrm{g/year}$ and $v\sim 500$~km.s$^{-1}$. Using these numbers we deduce that solar wind pressure is about 200 times weaker than solar radiation pressure. 

Because of this weakness it may not be important at this point to be able to determine the uncertainty in the solar wind albedo and the equivalent of $\eta$. It is certainly not as important as finding the truly stochastic part of solar wind, which is constituted by the coronal mass ejections hitting the satellite. Coronal mass ejections are violent outbursts from the solar surface, occurring every few days and carrying away masses $10^{14}$ to $10^{16}$~g at speeds up to 5000~km.s$^{-1}$~\citep{2010ApJ...711...75L,2010arXiv1002.3953Z}. We assume that the coronal mass flow is spread evenly in a spherical shell moving away from the Sun\footnote{This is certainly not a realistic assumption with respect to each single event, but may not be so bad on long term average.}. When particles from this shell hit the satellite, it gains momentum $\delta p$. The displacement after one day is 
\be d_{cme}=\dfrac{\delta p}{m} \cdot 1\ \mathrm{day} \ee
Using the largest numbers quoted, we obtain an estimate $d_{cme}\sim 0.0003\ \mathrm{m}$ for the purely stochastic part of solar wind displacement 
a day after the coronal mass ejection. 

\paragraph{Interplanetary dust} A relatively small non-gravitational perturbation, but completely stochastic in nature is produced by collisions of satellites with interplanetary dust. Our estimates of these perturbations are based on data from \citet{2006Icar..181..107N}, who  
studied the dust bands Karin and Veritas. The authors claim that these dust bands contribute from 30 to 50\% of all interplanetary dust in the near Solar system. They contribute $15000$ to $20000$ tons per year in dust accretion rate to the whole Earth. We assume that in the vicinity of the Earth the dust accretion is approximately isotropic and that it is proportional to the accreting area ($4\pi R^2=4{\cal A}=2{\mathrm m^2}$). Taking $36000$ tons per year as the total mass accretion rate on the Earth, we thus estimate the dust accretion rate per satellite to $1.4\times 10^{-4}$~g/year. 

We come to the following main conclusions: solar system dust presents a very mild drag resistance to satellite motion around the Earth. The typical orbital decay time scale is of the order of $\sim 3\cdot 10^{9}$years. The stochastic component is contributed mostly by collisions with 100-200~$\mu$m dust particles which move with respect to the Earth (and with respect to satellites) with an average velocity $\sim 17$~km.s$^{-1}$ and occur with a probability of less than one per year. The velocity change due to such a collision is $\delta v_{\mathrm dust}\sim 5-20\cdot 10^{-8}$~m.s$^{-1}$. The displacement of a satellite due to such a collision observed after one day would amount to $d_{\mathrm dust}\simeq 5-20$~mm. Note that such collisions could be detected and their impact measured if the satellites would monitor their mutual positions, since the probability for more than one collision to occur during the same day is very small.

\paragraph{Comparison} It is difficult to compare the strength of these so different non-gravitational perturbations without more extensive data on their character, changing strength and without carefully considering statistical properties of these noise sources. Such a study would go beyond the scope of this article and may not even be very useful before it could be verified  experimentally. To make a very rough comparison of importance of each of the listed perturbations, we assign to each perturbation a ``force'' and a displacement/day. As the worst case we simply use the full perturbing force to calculate the displacement of a satellite after one day. This is a gross overestimate in the case of solar radiation pressure or solar wind pressure. We attempt to refine such noise upper limit by estimating only the stochastic part of 1-day position error, i.e. the error remaining after the known part of the perturbing force has been taken into account with the best available data. Dust collision differs from all other perturbations considering their very low probability rate. Thus their displacement per day is just a very long term average. Our estimates are given in Table~\ref{PertForces}. The reader should be aware of the large uncertainty involved in these estimates.
\begin{table}[t]
\caption{Non-gravitational noise sources} \centering \begin{tabular}{c c c c
} \hline \hline perturbation &magnitude &displacement/day &stochastic
component\\ \hline clock noise &1ns/day & - &0.3m\\ radiation pressure &
4.3$\times 10^{-6}\mathrm{Nm^{-2}}$ &13m &$<$0.001m?\\ solar wind pressure
&2.2$\times 10^{-9}\mathrm{Nm^{-2}}$ &0.007m &$\sim 0.0003$m?\\ dust
collisions &$\sim 100\mu$m part & - &$\sim$ 0.00005m\\[1ex] \hline
\end{tabular} 
\label{PertForces}
\end{table}

Even if the above estimates are very rough it is clear that timing noise is by far the most important noise source that must be controlled on a longer time scale. However, it is worth pointing out that the non-gravitational perturbations, apart from the slow drag mentioned above, do not systematically change the orbital momentum of the satellites (after all the precessions due to gravitational perturbations have been properly taken into account). It implies that the area swept by the radius vector from the center of the Earth to each satellite can be a clock with a very good long term phase stability, limited in principle only by the uncertainty in the swept area. If each satellite was capable of detecting the timing signal of all the other visible satellites, it would be possible to use the timing capability of the constellation of satellites to determine its own space-time position. 
%
%
\section{Conclusion}
\label{s:conc}
We studied the use of null coordinates. They are user independent and realized by radio wave communications from a constellation of four satellites to a GNSS user. We introduced a local Schwarzschild coordinate system with three spatial orthonormal directions and one time-oriented vector, with coordinates $X,Y,Z,t$. Before taking into account the gravitational perturbations due to the Moon, Sun, planets, obliquity and rotation of the Earth, etc$\dots$, this local coordinate system can be considered as a global inertial system, i.e. oriented in fixed directions with respect to distant
stars. Since the main effect of gravitational perturbations is to make the local Schwarzschild frame precess about different axes determined by the
orbits of the perturbing bodies, and these precessions are very slow, the local Schwarzschild coordinate frame can to a high degree of accuracy be considered as inertial. Due to this fact it is possible to decouple the problem of the motion in the local Schwarzschild frame from the problem of connecting the local Schwarzschild frame to the global inertial frame. This second problem is well understood in the framework of classical non-relativistic gravitational perturbation theory, and will be the subject of a following article. 

In this article we concentrated on finding algorithms to describe the dynamics of bodies in the local Schwarzschild frame in full general relativity, in defining null coordinates in space-time that are tied to the constellation of satellites and in reading these coordinates in order to determine the Schwarzschild coordinates of an event in space-time. We obtained the analytic solutions for light-like (scattering) and time-like (closed) geodesics. We defined two algorithms: (i) an algorithm that calculates null coordinates $(\tau_1,\tau_2,\tau_3,\tau_4)$ corresponding to the local Schwarzschild coordinates $(t_o,x_o,y_o,z_o)$ of a user, and (ii) the ``reverse'' algorithm that calculates space-time coordinates $(t_o,x_o,y_o,z_o)$ of a user from its null-coordinates $(\tau_1,\tau_2,\tau_3,\tau_4)$. We assume that orbital parameters of satellites are known. 

The two algorithms have been written in Fortran 90 and checked against each other for consistency. All codes have been optimized for weak gravitational fields, and tested on a PC with a quad-core CPU and a laptop with a dual-core CPU. It takes $\sim 60-70$~ms to determine all four coordinates with 25 -- 32-digit accuracy. This makes our codes feasible and could be used in modern positioning devices. The effects of non-gravitational perturbations have been studied. We focused mainly on the stochastic part of these perturbations. Clock noise has been identified as the most important contributor affecting the accuracy of position determination. By magnitude the second perturbing force is solar radiation pressure. We note, however, that this force is very stable and could, at least in principle, be modeled to a very high precision. Therefore, its stochastic effect is not expected to contribute significantly more to position noise than the remaining two truly stochastic perturbations: solar wind pressure and collisions with interplanetary dust.

We believe to have shown that the use of a fully relativistic code in GNSS systems offers an interesting alternative to the use of post-Newtonian approximations, and presents no technical obstacle. The advantages of using a fully relativistic code over the use of classical post-newtonian codes go well beyond aesthetics. The stochastic part of the non-gravitational perturbation forces has no preferred direction; therefore they are not expected to change average values of orbital angular momenta of satellites. It implies that area velocities of satellites are good constants of motion perturbed only by deterministic gravitational perturbations. Therefore we suggest that the area covered by the radius vector from the center of the Earth to the satellites be used as a stable measure of proper time, in particular to correct the long term clock drifts. The constellation would have a greater autonomy in defining its own local Schwarzschild frame in which the predictability of dynamics can be taken advantage of.

This concept of autonomy offers an increased accuracy and long term stability to the global positioning system. It promises us a tool for a systematic study of the space-time metric around the Earth and a deeper insight into the Earth internal dynamics. In order to realize such an autonomous constellation, it must be self-consistent. By this we mean that if satellites determine their orbits by using positioning data from other satellites in the constellation, they should obtain the same constants of motion that were used in the definition of the global positioning system. Such self-consistency at the maximum level of precision is not obvious, but can be reached if each satellite is also a receiver of timing signals of all other satellites. We believe we can design a mathematical procedure to drive any initial GNSS solution toward the best self consistent solution.

\end{document}